\documentclass[conference]{IEEEtran}
\usepackage{cite,amsmath,amssymb,amsfonts,algorithmic,multirow,graphicx,textcomp,xcolor,booktabs,listings,url}
\usepackage[ruled,linesnumbered]{algorithm2e}
\usepackage{subfigure}
\begin{document}
\title{Resource-efficient Parallel Split Learning in Heterogeneous Edge Computing}

\author{
Mingjin~Zhang,~Jiannong~Cao,~Yuvraj~Sahni,~Xiangchun~Chen,~Shan~Jiang\\
The Hong Kong Polytechnic University \\ 
\{csmzhang, csjcao, csxcchen\}@comp.polyu.edu.hk, \{cs-shan.jiang, yuvraj-comp.sahni\}@polyu.edu.hk
}

\IEEEtitleabstractindextext{%
\begin{abstract}
Edge AI has been recently proposed to facilitate the training and deployment of Deep Neural Network (DNN) models in proximity to the sources of data. To enable the training of large models on resource-constraint edge devices and protect data privacy, parallel split learning is becoming a practical and popular approach. However, current parallel split learning neglects the resource heterogeneity of edge devices, which may lead to the straggler issue. In this paper, we propose EdgeSplit, a novel parallel split learning framework to better accelerate distributed model training on heterogeneous and resource-constraint edge devices. EdgeSplit enhances the efficiency of model training on less powerful edge devices by adaptively segmenting the model into varying depths. Our approach focuses on reducing total training time by formulating and solving a task scheduling problem, which determines the most efficient model partition points and bandwidth allocation for each device. We employ a straightforward yet effective alternating algorithm for this purpose. Comprehensive tests conducted with a range of DNN models and datasets demonstrate that EdgeSplit not only facilitates the training of large models on resource-restricted edge devices but also surpasses existing baselines in performance.
\end{abstract}

}

\maketitle

\IEEEdisplaynontitleabstractindextext

%
\IEEEpeerreviewmaketitle

\section{Introduction}\label{sec:introduction}
AI models have traditionally been trained and deployed in centralized cloud environments due to their intensive resource requirements, using data gathered from various end-user devices. Despite its popularity, it often encounters issues such as high communication costs, delayed responses, and privacy risks \cite{zhang2022eaas}. To address these challenges, a new paradigm known as Edge AI has emerged, emphasizing the training and implementation of AI models on edge devices (like edge servers, gateways, and smartphones), which are situated closer to where the data originates \cite{chen2019deep}.

A crucial challenge within Edge AI is the development of accurate models that learn quickly from distributed data across numerous edge devices. Federated Learning (FL) \cite{mcmahan2017communication}, a leading approach in this context, facilitates collaborative model training among a variety of edge devices while safeguarding user privacy. In FL, edge devices locally train the AI model required by the FL server and only send model updates, such as weights or gradients, back to the server for integration. The server then dispatches the combined parameters for the next training cycle. This iterative process continues until the model reaches satisfactory accuracy levels. FL has proven effective in several applications \cite{rieke2020future,long2020federated}.

To enhance the performance of edge AI applications, it's often beneficial to increase the parameters of Deep Learning (DL) models. However, the training of large models poses a challenge for resource-limited workers, primarily due to their restricted CPU and memory capabilities. Hence, split learning \cite{gupta2018distributed, poirot2019split} was proposed to enable the model training on low-resource mobile devices by splitting the full model between server and clients. SL partitions a full model into two partial models, i.e., client model and server model. The client model is trained on edge devices, while the server model is trained on the server with the representation of the partition layer (also known as activations) transmitted between the server and the clients. Since these representations are considerably smaller than the full model, the burden on communication is also significantly reduced. 

However, split learning is a sequential training paradigm. Edge devices take turns to collaboratively train with the server. It is not suitable for parallel federated learning. Recently, parallel split learning \cite{thapa2020splitfed, jeon2020privacy, han2021accelerating} was proposed to combine the benefits of both federated learning and split learning. SplitFed \cite{thapa2020splitfed} is the first attempt to enable the parallel training of the client-side model. \cite{jeon2020privacy} adaptively adjust the mini-batch size considering the amount of local data on each edge device. \cite{han2021accelerating} introduce an auxiliary network to avoid frequent intermediate data transmission. However, current parallel split learning seldom considers the resource heterogeneity of edge devices. There are various edge devices, such as mobile phones, edge servers, and Raspberry Pi. The computing and networking capabilities of those devices are usually vastly different. In each training round, the FL server is required to wait for the updated parameters of all the participants, which may lead to long waiting time due to device heterogeneity.

In this paper, we propose EdgeSplit, a novel parallel split learning framework, aiming at enhancing the efficiency of federated learning on heterogeneous and resource-constraint edge devices. EdgeSplit is designed to optimize resource utilization and orchestrate the training tasks between edge devices and the FL server. More specifically, a full model is partitioned into a set of shallow models with different depths, adapting to the heterogeneous computation resources of edge devices. Each device trains a segment of the full model, while offloading a portion of the training task to the FL server. The server plays a dual role: completing the remaining training tasks and aggregating the model parameters.

However, it is non-trivial to decide the optimal model partition point for each edge device. DNN models are characterized by significant variations in computational workload and parameter sizes across their different layers. Simple model partitioning approach is often ineffective in decreasing the overall end-to-end latency, as it fails to account for the complex, layered nature of DNN models and the unique demands of each layer. Hence, to reduce the training time, we mathematically formulate a training task scheduling problem to decide how to split the training model for each edge device and the bandwidth between an edge device and the server, in terms of both edge devices computing capabilities and network bandwidth. We propose an efficient alternating algorithm to solve the problem with the objective of minimizing the overall training time. The algorithm iteratively optimizes model splitting and bandwidth allocation strategy and can quickly reach the optimal objective. 

We develop a real-world testbed with both physical and virtual edge devices to emulate the system in large scale. We have comprehensively evaluated the performance of EdgeSplit under various DNN models and datasets. The evaluation results show that our proposed EdgeSplit can achieve up to $5.5$x training speed improvement. Our contributions are summarized as follows:
\begin{itemize}
    \item We propose a novel parallel split learning framework to adaptively split the model among edge devices and the FL server in heterogeneous and resource-constraint edge computing environments and achieving training acceleration without compromising accuracy.
    \item We formulate a training task scheduling problem by jointly considering the neural network splitting and bandwidth allocation with the objective of reducing the model training time. 
    \item We propose an efficient alternating algorithm to solve the problem and evaluate the performance of EdgeSplit on a real-world testbed with various benchmark models and datasets. The results indicate the superiority of EdgeSplit. 
\end{itemize} 

\section{Framework of EdgeSplit}\label{sec: framework}

\begin{figure}[t]
	\includegraphics[width=0.7\linewidth]{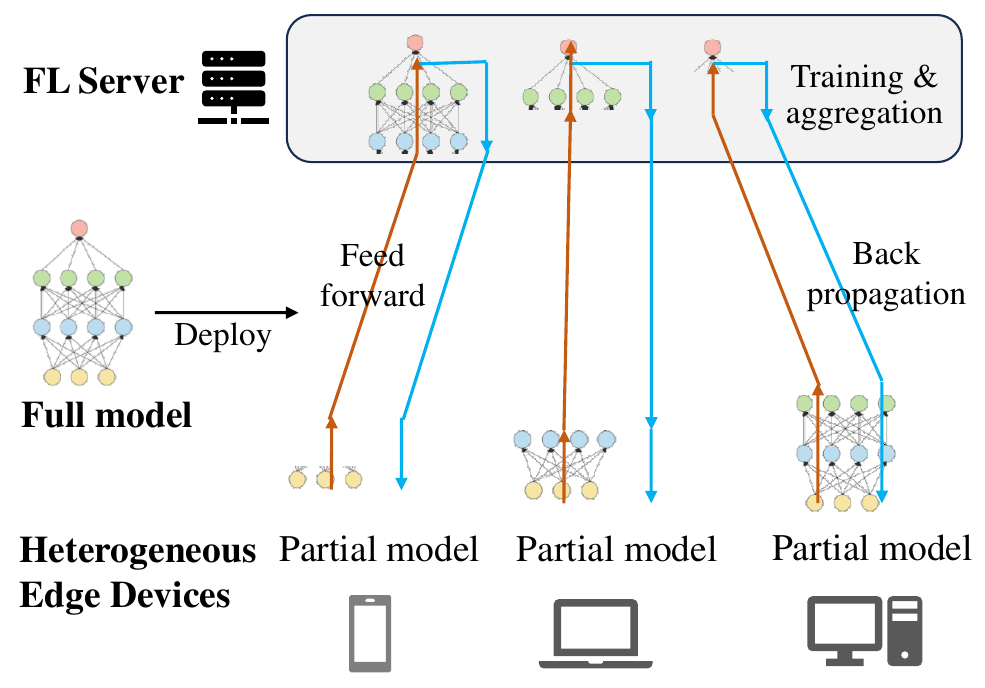} 
	\centering
	\caption{Framework of EdgeSplit. Edge devices train part of the full model with different depths adapting to local resources and offload the rest of model training task to the FL server for acceleration.}
	\label{EdgeSplit}
\end{figure}

Fig.~\ref{EdgeSplit} shows the overview of the parallel split learning framework. EdgeSplit first decides the partition points and bandwidth for each edge device, according to the heterogeneous computation capabilities, the bandwidth, and the characteristics of the DNN models, such as the size of output for each layer. The details are shown in Sec.~\ref{sec: optimization}. 

After getting the best partition points, EdgeSplit will perform online split training. The training process consists of the following steps.
\begin{itemize}
    \item \textbf{Step 1}: Each edge device performs feed training with localized data and shares the outputs/activations of the partial models with the server
    \item \textbf{Step 2}: The server performs the rest of the feed training and back propagation and sends back the gradients to edge devices
    \item \textbf{Step 3}: Each edge device gets the gradients and does local back propagation and updates its weights
    \item \textbf{Step 4}: Repeat the following steps according to the number of batches
    \item \textbf{Step 5}: Each edge device sends the updated weights to the server for aggregation
    \item \textbf{Step 6}: The server aggregates all updated weights from edge devices and sends back the aggregated weights to edge devices
\end{itemize}

\textbf{Benefits.} EdgeSplit is suitable for accelerating federated learning on heterogeneous edge devices for three reasons. 
\begin{itemize}
    \item \textbf{Resource-efficient.} Edge devices only need to train part of the full model, subjecting to the local computing capabilities and the bandwidth between an edge node and the server. It enables training resource-greedy AI models on resource-constraint edge devices, such as raspberry pi, which has only 1GB memory.
    \item \textbf{Communication-efficient.} In EdgeSplit, only activations of the partition layer and partial of full weights are transmitted between edge devices and the server, which reduces network pressure and achieves fast training.
    \item \textbf{No Accuracy Loss.} EdgeSplit only offloads partial training tasks to the FL server. It does not compress the data or modify any hyper-parameters of the training. There is no accuracy loss of EdgeSplit compared to FL. 
\end{itemize}

\section{Model Splitting and Bandwidth Allocation}\label{sec: optimization}
This section shows details about the task scheduling to decide the best model partition points for each edge device. 

\textbf{System Model.} We consider a network consisting of $M$ edge devices and a server. The edge devices have heterogeneous computation capabilities, and the server is located in the remote cloud and is much more powerful than edge devices in terms of computation capability. The edge devices and the server are interconnected, and the total bandwidth between the server and edge devices is denoted by $B$. Bandwidth between an edge device $i$ and the server is $B_{i}$, $ 1 \leq i \leq M$. We assume the DNN model is with $N$ feasible partition layers. $O_{j}$ represents the size of output/activations of layer $j, 1 \leq i \leq N$. $P_{j}$ is number of parameters from layer 1 to layer $j$. 

\textbf{Problem Formulation.}
The forward computation time $T_{batch}^{f}$, the back propagation time $T_{batch}^{b}$, and communication time for edge device $i$ in a batch is calculated by the following equations:
\begin{equation}\label{eq: forward}
     T_{batch}^{f} = \sum_{j=1}^{N} T_{i,j}^{f}*X_{i,j} + \sum_{j=1}^{N} S_{i,j}^{f}*X_{i,j}
\end{equation}
\begin{equation}
\begin{split}
    T_{batch}^{b} = \sum_{j=1}^{N} (S_{i,j}^{b} + T_{i,j}^{b}) *X_{i,j}
\end{split}
\end{equation}
\begin{equation}
     T_{batch}^{g} = \frac{\sum_{j=1}^{N} O{j}*X_{i,j}}{B_{i}}*2
\end{equation}

\noindent where $X_{i,j}$ is a binary variable. $X_{i,j}$ equals to $1$ if layer $j$ of the model is selected as the partition point for edge device $i$. Otherwise, $X_{i,j}$ equals to $0$. $T_{i,j}^{f}$ is the forward time from layer $1$ to layer $j$ on device $i$. $S_{i,j}^{f}$ is the forward time from layer $j+1$ to end layer on the server. $T_{i,j}^{b}$ is the backward time from layer $j$ to layer 1 on device $i$. $S_{i,j}^{b}$ is the backward time from end layer to layer $j+1$ on the server. Hence, $\sum_{j=1}^{N} T_{i,j}^{f}*X_{i,j}$ represents the feed forward time on local edge device $i$, and $\sum_{j=1}^{N} S_{i,j}^{f}*X_{i,j}$ represents the rest of feed forward time performed on the server. $\sum_{j=1}^{N} O{j}*X_{i,j}$ indicates the amount of activations/gradients for device $i$. The activations and gradients are assumed to have the same size, as they are decided by the shape of the partition layer, i.e., the number of neurons in this layer. 

Note that edge device $i$ also has to receive the initial weights from the server at the beginning of the training and send back the updated weights when finishing a round of training. The communication time to receive and send the weights is calculated by Eq.~(\ref{eq:weights}), where $P{j}$ is the number of parameters from layer 1 to layer $j$.
\begin{equation}\label{eq:weights}
     T_{batch}^{w} = \frac{\sum_{j=1}^{N} P{j}*X_{i,j}}{B_{i}}*2
\end{equation}

The time for a round of training of edge device $i$ is shown in Eq.~\ref{eq:one-round}, where $b$ is the number of batches.

\begin{equation}\label{eq:one-round}
     T_{i}^{r} = b*(T_{batch}^{f} + T_{batch}^{b} + T_{batch}^{g}) + T_{batch}^{w}
\end{equation}

\textbf{Objective.} The problem of minimizing the training time of EdgeSplit is formulated as follows. We denote this problem as $P1$. As shown in the objective function Eq.~(\ref{eq: objective}), the overall training time is determined by the maximum training time of edge devices, and our objective is to minimize the maximum training time for acceleration.  
\begin{equation}\label{eq: objective}
     \textbf{P1:} \quad \min_{X_{i,j}, B_{i}} \max \left\{T_{1}^{r}, T_{2}^{r}, ..., T_{i}^{r}\right\}_{i=1}^{M} 
\end{equation}

\begin{equation}\label{eq: partition}
     \sum_{j=1}^{N} X_{i,j} = 1, \quad \forall i
\end{equation}
\begin{equation}\label{eq: bandwidth}
    \sum_{i=1}^{M} B_{i} \leq B
\end{equation} 

\textbf{Problem Solution.}
There are both binary variable $X_{i,j}$ and continuous variable $B_{i}$ in $P1$. The problem is a mixed integer non-linear problem, which is proven to be NP-hard in literature and is hard to solve. To solve the problem, we first simplify the original problem $P1$ and then propose an efficient alternating algorithm to solve it. $T_{i}^{r}$ can be simplified and rewritten as follows:
\begin{equation}\label{eq: rewritten}
     T_{i}^{r} = \sum_{j=1}^{N} (A_{i,j}+\frac{C_j}{B_i})*X_{i,j} 
\end{equation}

\noindent where $ A_{i,j} = b*(T_{i,j}^{f} + S_{i,j}^{f} + S_{i,j}^{b} + T_{i,j}^{b})$ and $C_{j} = (b*O_{j} + P_{j})*2$. In Eq.~(\ref{eq: rewritten}), $A_{i,j}$ and $C_{j}$ are deterministic and there are two variables $X_{i,j}$ and $B_{i}$. We observe that once fix $B_{i}$, there is an analytical solution for $X_{i,j}$, indicated by Eq.~\ref{eq: Xij}, and once fix $X_{i,j}$, the original problem $P1$ becomes a convex problem $P2$, indicating by Eq.~\ref{eq: Bi}. Instead of relaxing $X_{i,j}$ to a continuous variable, which is usually adopted, we propose an alternative minimization method to solve the problem, which alternatively searches and optimizes $X_{i,j}$ and $B_{i}$. There are three steps of the method. 

\begin{equation}\label{eq: Xij}
 X_{i,j} = \begin{cases}
 1, \quad &j = \mathop{\arg\min}\limits_{j}(A_{i,j}+\frac{C_j}{B_i})\\
 0 &\text{otherwise.}
 \end{cases}
\end{equation} 

\begin{equation}\label{eq: Bi}
\begin{aligned}
\textbf{P2:} \quad \min_{B_{i}} \max \left\{T_{1}^{r}, T_{2}^{r}, ..., T_{i}^{r}\right\}_{i=1}^{M} \\
\end{aligned}
\end{equation}

The algorithm is shown in Alg.~\ref{algo:solution}. 

\begin{algorithm}[t]
\caption{Joint model partition and bandwidth allocation}\label{algo:solution}
\KwIn{Profiled data $T_{i,j}^{f}$, $S_{i,j}^{f}$, $ S_{i,j}^{b}$, $T_{i,j}^{b}$, $O_{j}$, $P_{j}$; total bandwidth $B$; number of batches $b$ }
\KwOut{the model splitting strategy $X_{i,j}$ and bandwidth allocation strategy $B_{i}$ }
Initialize $B_{i}$ for all edge devices\;
Initialize a large one-round training time $T_{opt} \gets INF$\;
\For{iterations}{
    \tcp{Step 1}
    Calculate $X_{i,j^{*}}^{*}$ by solving Eq.~(\ref{eq: Xij})\;
    Fix model splitting strategy by assigning $X_{i,j^{*}}^{*} \gets 1$\;
    \tcp{Step 2}
    Solve convex problem $P2$ and get optimal objective value $T_{i^{*}}^{r^{*}}$ and $B_{i}^{*}$ for all edge devices\;
    \uIf{$T_{opt} > T_{i^{*}}^{r^{*}}$}{
        $T_{opt} \gets T_{i^{*}}^{r^{*}}$\;
        Fix $B_{i}^{*}$ for all edge devices\;
    }
    \Else{
        break\;
    }
    
}
\Return $X_{i,j^{*}}^{*}$, $B_{i}^{*}$
\end{algorithm}

\section{Experimental Evaluation}\label{sec: exp}

\subsection{Experimental Setup}
We use Nvidia Jetson Xavier AGX to represent physical devices. Xavier AGX is with $32$GB memory and 12 cores. For the FL server, we employ a robust server equipped with $64$GB memory to manage the edge devices and the training process. To simulate virtual edge devices, we utilize an edge server capable of hosting Docker containers emulating virtual edge devices. The edge server is with $192$GB Memory and $128$ cores CPUs. Docker's flexibility allows us to easily adjust the number of CPU cores and the memory allocated to each container, effectively varying their computational power. Consequently, a container with a greater number of CPU cores exhibits enhanced computational capabilities.

\subsection{Datasets and Models}
We conduct tests using various DNN models, including LeNet, VGG-16, ResNet-50, and ResNet-101. They are classical and representative models and are extensively used in both academia and industry. Moreover, they have a variant depth of neural network from $5$ to $101$ layers, with varying sizes of parameters and computation workloads. We use three standard datasets, i.e., Fashion-MNIST, MNIST, and CIFAR-10.

\subsection{Baselines and Metrics}
We compare EdgeSplit with the following three baselines. 1) FedAvg \cite{mcmahan2017communication}. Edge devices perform the local training of the full model, and the server only conducts weights aggregation. The bandwidth is equally shared among edge devices. 2) Adaptive FL. Adaptive bandwidth allocation for accelerating FL is used in \cite{ren2020accelerating,wang2019adaptive}. In this case, model training is also done locally. However, the bandwidth allocation is decided by solving problem P2 in \textsection~\ref{sec: optimization}. 3) SplitFed \cite{thapa2020splitfed}. Partition points for all the edge devices are identical. We allocate half of the layers on the edge devices and half of them on the server.

\subsection{Results and Analysis}
In our experiments, we evaluate the performance of EdgeSplit and compare it with other baseline methods under a variety of conditions. Throughout these tests, we maintain a client selection ratio of 1, implying that all edge devices are involved in the training process. Additionally, we set the batch size to 128 and keep the number of local epochs at 1.

\begin{table*}[t]
    \centering
    \caption{Comparison of per-round training time. Best partition points and acceleration ratio to vanilla FL are given.}
    \label{t:one-round-2} 
     \begin{tabular}{c|c|c|cc|ccc}
        \toprule
        Model& FedAvg & Adaptive FL & SplitFed & Acceleration & EdgeSplit & Best Partition Points & Acceleration \\
        \midrule
        LeNet & $237.9$ & $234.2$ & $164.3$ & $1.4$x & $134.8$ & [1,1,1,1,2,2,4,4] & 1.76x  \\ 
        VGG16 & 969.6 & 953.2 & 384.2 & 2.5x & $243.4$ & [1,1,3,3,13,13,13,13] & 3.9x \\  
        ResNet50 & 1806 & 1800.4 & 696.6 & 3.0x & $330.2$ & [1,1,1,1,2,2,49,49] & 5.5x \\
        ResNet101 & 1355.7 & 1352.3 & 693.6 & 1.9x & $308.1$ & [1,1,1,1,1,1,8,8] & 4.4x \\
        \bottomrule 
     \end{tabular}
\end{table*}

A hybrid testbed is utilized to simulate the diversity of heterogeneous edge devices, enabling a quantitative analysis of EdgeSplit's performance. For this purpose, we configure the memory of the emulated devices (represented by Docker containers) to be 12GB, which is suitable for training larger models, e.g., ResNet50 and ResNet101. To further emulate a range of virtual edge devices with varying computational powers, we set up these containers with different CPU core configurations: 1 core, 3 cores, and 5 cores. The number of four types of edge devices is identical. 

\begin{figure*}[t]
\centering
\subfigure[LeNet on MNIST]{
\includegraphics[width=0.244\linewidth]{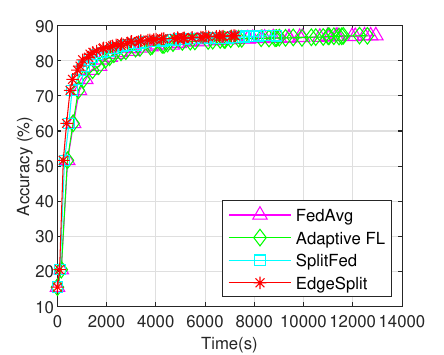}
}\hspace{-3mm}
\subfigure[VGG16 on CIFAR-10]{
\includegraphics[width=0.244\linewidth]{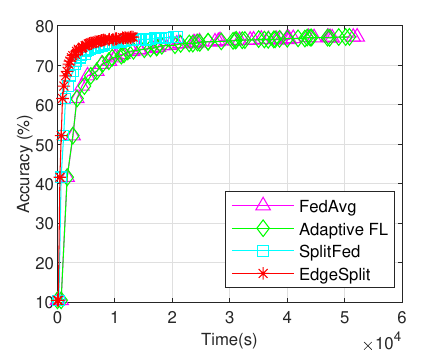}
}\hspace{-3mm}
\subfigure[ResNet50 on FashionMNIST]{
\includegraphics[width=0.244\linewidth]{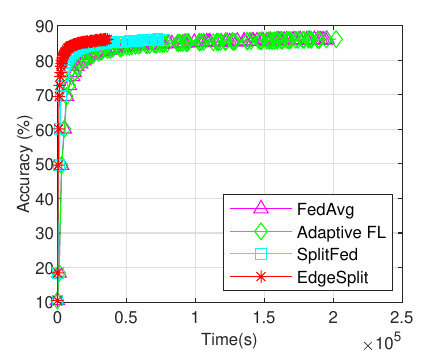}
}\hspace{-3mm}
\subfigure[ResNet101 on CIFAT-10]{
\includegraphics[width=0.244\linewidth]{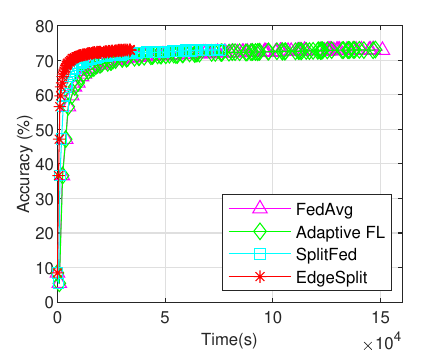}
}
\caption{Convergence Time v.s. Accuracy. EdgeSplit achieves fast convergence without accuracy loss}
\label{exp-convergence}
\end{figure*}

Table~\ref{t:one-round-2} presents the training times per round involving eight edge devices operating under a 30Mbps bandwidth. It's evident that model partitioning methods, namely SplitFed and EdgeSplit, significantly enhance training speeds. This improvement is attributed to the model split, which not only shortens local training duration but also delegates a portion of computational tasks to the more capable FL server, thus speeding up the process. Notably, EdgeSplit surpasses SplitFed in performance, with up to a 5.5x increase in training speed for ResNet50. This is due to EdgeSplit's ability to adaptively select partition points according to the varying resource capacities of each edge device. We also note that Adaptive FL exhibits performance akin to Vanilla FL in this scenario. This similarity in performance is likely because the available bandwidth is relatively sufficient, making computation the primary limiting factor. Furthermore, Figure~\ref{exp-convergence} illustrates the relationship between training time and accuracy. Here, EdgeSplit demonstrates a reduced training time per round and achieves rapid convergence without compromising accuracy.

\section{Related Work}\label{sec: literature}

\subsection{Efficient Federated Learning}
Previous works usually focused on reducing the total amount of transmission bits to improve communication efficiency. Sparsification \cite{konevcny2018randomized} and quantization \cite{alistarh2017qsgd} are two notable methods. The former selects only a fraction of the parameters (e.g., gradient or weights) to be sent to the FL server. The latter aims to represent the model update with fewer bits, e.g., using 8-bit or 16-bit low-precision representation. Other works consider heterogeneous computation and networking resources to improve efficiency. \cite{ren2020accelerating} jointly optimize the batch size selection and communication resource allocation in wireless federated edge learning systems. \cite{wang2019adaptive} and \cite{yang2022optimizing, yang2021tree} optimize the aggregation frequency considering heterogeneous local edge resources. \cite{chen2020joint} formulates a joint user selection and resource allocation policy under limited bandwidth to minimize the training loss. However, those works neglect the resource constraint of edge devices. Some edge devices with limited memory may not have the ability to burden the training tasks due to memory constraint.

\subsection{Split Learning}
Model spliting is previously used for collaborative model inference \cite{zhang2022ents,zhang2022blockchain}. Recently, some works also use model splitting to train the model in edge computing environments. \cite{gupta2018distributed, poirot2019split} proposed split learning, which partitions the deep neural network into two parts, where the shallow part is trained on the client, and the deep part is trained on the server. Such schema can leverage the heterogeneous computing capacity of the clients and server to enable large model training on low-resource mobile devices. However, the training is performed in a sequential manner. SplitFed \cite{thapa2020splitfed} is the first attempt to integrate splitting learning and federated learning to support parallel and distributed model training. \cite{jeon2020privacy} fixes the first layer of the model on clients and adaptively adjusts the local batch size catering to the amount of local data. LocSplitFed \cite{han2021accelerating} leverage local loss function to avoid model aggregation in each round for reducing the amount of data transmission. However, they do not consider how to split the model for faster convergence with respect to the heterogeneous computation resources, and they do not consider the networking model in practical federated learning systems. 

\section{Conclusion}\label{sec: conclusion}
In this work, we propose EdgeSplit to expedite federated learning on heterogeneous and resource-limited edge devices. By segmenting a full DNN model adapting to the heterogeneous resources of edge devices, EdgeSplit facilitates the training of large-scale models on devices with limited resources. It offloads a portion of the training workload to the more powerful FL server, thereby significantly reducing the total training duration. Experiments under various settings show the performance of EdgeSplit surpasses the baselines. EdgeSplit can achieve notable acceleration by jointly deciding the optimal model partition point and bandwidth allocation for each edge device.

\ifCLASSOPTIONcaptionsoff
  \newpage
\fi

\bibliographystyle{IEEEtran}
\bibliography{ref.bib}

\end{document}